\documentclass[a4paper]{jpconf}
\usepackage{graphicx}
\begin{document}
\title{Magnetic-Field Induced Gap in One-Dimensional Antiferromagnet KCuGaF$_6$}

\author{R. Morisaki$^1$, T. Ono$^1$, H. Tanaka$^1$ and H. Uekusa$^2$}

\address{$^1$Department of Physics, Tokyo Institute of Technology, Oh-okayama, Meguro-ku, Tokyo 152-8551, Japan}
\address{$^2$Department of Chemistry, Tokyo Institute of Technology, Oh-okayama, Meguro-ku, Tokyo 152-8551, Japan}

\ead{morisaki@lee.phys.titech.ac.jp}

\begin{abstract}
Magnetic susceptibility and specific heat measurements in magnetic fields were performed on an $S=1/2$ one-dimensional antiferromagnet KCuGaF$_6$. Exchange interaction was evaluated as $J/k_{\rm B}\simeq 100$ K. However, no magnetic ordering was observed down to 0.46 K. It was found that an applied magnetic field induces a staggered magnetic susceptibility obeying the Curie law and an excitation gap, both of which should be attributed to the antisymmetric interaction of the Dzyaloshinsky-Moriya type and/or the staggered $g$-tensor. With increasing magnetic field $H$, the gap increases almost in proportion to $H^{2/3}$. 
\end{abstract}

\section{Introduction}
It is known that magnetic excitations in an $S=1/2$ antiferromagnetic Heisenberg chain (AFHC) represented by the Hamiltonian ${\cal H}=\sum_iJ\vec S_i\cdot \vec S_{i+1}$ are gapless \cite{dCP}. The excitations are gapless even in the external magnetic field, although the wave vector $q$ for the gapless excitation varies with the expectation value of spin $\langle S^z\rangle$ as $q=2{\pi}\langle S^z\rangle$ \cite{Ishimura}. Recently, unexpected excitation gaps induced by the external magnetic field have been observed in some $S=1/2$ AFHC systems, Cu(C$_6$H$_5$COO)$_2$$\cdot$3H$_2$O \cite{Dender,Asano}, Yb$_4$As$_3$ \cite{Oshikawa1}, [PM$\cdot$Cu(NO$_3$)$_2\cdot$(H$_2$O)$_2$]$_n$ (PM=pyrimidine) \cite{Feyerherm,Zvyagin} and CuCl$_2$$\cdot$2((CH$_3$)$_2$SO) \cite{Kenzelmann}. The field-induced gaps ${\Delta}(H)$ in these compounds were commonly described as
\begin{equation} 
{\Delta}(H)=AH^{\alpha},
\end{equation}
with ${\alpha}\simeq 2/3$. Coefficient $A$ is strongly dependent on the field direction. 
The low symmetric crystal structures in these compounds allow the existence of the alternating $g$-tensor and the antisymmetric interaction of the Dzyaloshinsky-Moriya (DM) type, ${\cal H}_{\rm DM}=\sum_i\vec D_{i}\cdot \vec S_i\times \vec S_{i+1}$ with the alternating $\vec D$ vector, both of which can lead to the effective staggered field when an external magnetic field is applied. Taking these effects into consideration, Oshikawa and Affleck \cite{Oshikawa2,Affleck} investigated the following model:
\begin{equation}
\mathcal{H}=\sum_{i}\,\left[J\vec S_i\cdot \vec S_{i+1}-g{\mu}_{\rm B}HS_i^z+(-1)^ig{\mu}_{\rm B}hS_i^x\right], 
\end{equation}
where $h=c_{\rm s}H$ is the effective staggered field induced by the applied field. Mapping this model onto the quantum Sine-Gordon model, they derived 
\begin{equation}
{\Delta}(H)\simeq 1.8J\left(\frac{g{\mu}_{\rm B}h}{J}\right)^{2/3}\left\{\ln \left(\frac{J}{g{\mu}_{\rm B}h}\right)\right\}^{1/6}.
\end{equation}
Since the logarithmic correction in eq. (3) is not significant, their result explains the experimental results well.

For the comprehensive understanding of static and dynamic properties in the systems described by eq. (2), new compounds having different interaction constants are needed. In this paper, we show that a new $S=1/2$ AFHC system KCuGaF$_6$ exhibits the field-induced gap.
KCuGaF$_6$ has a monoclinic crystal structure with space group $P2_1/c$, in which Cu$^{2+}$ and Ga$^{3+}$ ions form a pyrochlore lattice \cite{Dahlke}. Since Cu$^{2+}$ ions are arranged almost straightforward and neighboring Ga$^{3+}$ ions are nonmagnetic, the exchange interaction between neighboring Cu$^{2+}$ ions has one-dimensional nature. In KCuGaF$_6$, CuF$_6$ octahedra are elongated perpendicular to the chain direction parallel to the $c$-axis due to the Jahn-Teller effect. Consequently, the hole orbitals $d(x^2-y^2)$ of Cu$^{2+}$ ions are linked along the chain direction through the $p$ orbitals of F$^-$ ions, which can generate strong antiferromagnetic exchange interaction along the chain. The elongated axes are alternate. This low symmetric crystal structure can lead to both the staggered $g$-tensor and the DM interaction. Therefore, the magnetic model for the present system should be expressed by eq. (2).

\section{Experimental}
Single crystals of KCuGaF$_6$ were grown by the vertical Bridgman method from the melt of an equimolar mixture of KF, CuF$_2$ and GaF$_3$ sealed in Pt-tube. Crystals obtained were identified to be KCuGaF$_6$ by X-ray powder and single crystal diffractions. Magnetic susceptibilities were measured by using a SQUID magnetometer (Quantum Design MPMS XL). Specific heat measurements were carried out down to 0.5K in magnetic fields of up to 9 T, using a Physical Property Measurement System (Quantum Design PPMS) by the relaxation method. 

\section{Results and Discussion}
Figure 1 shows the temperature dependences of the magnetic susceptibilities $\chi$ of KCuGaF$_6$ measured at $H=0.1$ T for $H\parallel c$, $H\perp(1,1,0)$ and $H\parallel [1,1,0]$. For $H\perp(1,1,0)$ and $H\parallel [1,1,0]$, the susceptibilities exhibit broad maxima at $T\sim 60$ K characteristic of one-dimensional antiferromagnets. With decreasing temperature, the susceptibilities increase rapidly below 20 K, obeying the Curie law. For $H\parallel c$, the Curie term is so large that the broad susceptibility maximum is hidden. The Curie term is intrinsic to the present system, because its magnitude depends on field direction and is independent of sample. The staggered susceptibility for AFHC has a tendency to diverge for $T\rightarrow 0$. Thus, we assume that the Curie term arises from the staggered field induced by the external field through the staggered $g$-tensor and the DM interaction.

According to Oshikawa and Affleck \cite{Oshikawa2,Affleck}, the magnetic susceptibility for the model (2) is expressed as 
\begin{equation}
{\chi}\simeq {\chi}_{\rm u}+c_{\rm s}^2{\chi}_{\rm s},
\end{equation}
where ${\chi}_{\rm u}$ is the uniform susceptibility for $S=1/2$ AFHC without anisotropy and has been calculated precisely by Eggert {\it et al.} \cite{Eggert}. ${\chi}_{\rm s}$ is the staggered susceptibility given by
\begin{equation}
{\chi}_{\rm s}\simeq \frac{0.278Ng^2\mu_{\rm B}^2}{4k_{\rm B}T}\left\{\ln\left(\frac{J}{k_{\mathrm{B}}T}\right)\right\}^{1/2},
\end{equation}
Fitting eq. (4) to the susceptibility data, we obtain $J/k_{\mathrm{B}}\simeq 100$ K and $c_{\rm s}\simeq 0.54, 0.29$ and $0.22$ for $H\parallel c$, $H\perp(1,1,0)$ and $H\parallel [1,1,0]$, respectively. The exchange constant in the present system is much larger than $J/k_{\mathrm{B}}\simeq 18$ K for Cu(C$_6$H$_5$COO)$_2$$\cdot$3H$_2$O \cite{Dender,Asano} and $J/k_{\mathrm{B}}\simeq 36$ K for [PM$\cdot$Cu(NO$_3$)$_2\cdot$(H$_2$O)$_2$]$_n$ (PM=pyrimidine) \cite{Feyerherm} The proportional coefficient $c_{\rm s}=h/H$ is strongly dependent on the field direction and is the largest for $H\parallel c$. Since the anisotropy of the $g$-factor is usually ${\Delta}g\sim 0.1$ in copper compounds, the large proportional coefficient $c_{\rm s}\simeq 0.54$ for $H\parallel c$ cannot be produced by the staggered $g$-tensor only. This indicates the existence of the strong DM interaction with the $\vec D$ vector perpendicular to the $c$-axis.
\begin{figure}[ht]
\begin{center}
\includegraphics[width=20pc]{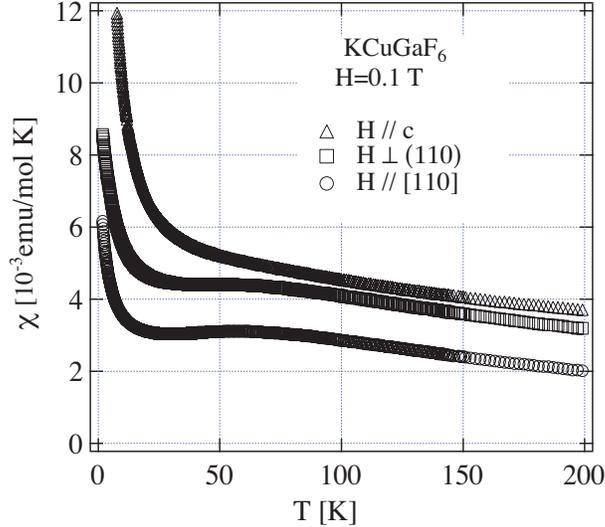}\hspace{2pc}%
\end{center}
\caption{Temperature dependences of magnetic susceptibilities $\chi$ in KCuGaF$_6$ measured at $H=0.1$ T for $H\parallel c$, $H\perp(1,1,0)$ and $H\parallel [1,1,0]$. The values of the susceptibilities are shifted upward consecutively by $1\times 10^{-3}$ emu/mol. \label{ai}}
\end{figure}

Figure 2 shows the low-temperature specific heat $C$ in KCuGaF$_6$ measured at various magnetic fields for $H\parallel c$. No magnetic ordering was observed down to 0.46 K, which is indicative of good one-dimensionality of the present system. At zero magnetic field, $C$ exhibits linear temperature dependence below 4 K where the lattice contribution is small. The low-temperature molar specific heat for the $S=1/2$ AFHC is given by $C=2RT/(3J/k_{\rm B})$ \cite{Klumper}. Using this equation, we obtain $J/k_{\rm B}\simeq 100$ K, which coincides with $J$ obtained from susceptibility data.

With increasing applied field, $C$ decreases exponentially toward zero. This behavior is characteristic of gapped ground state. We analyze low-temperature specific heat for $H\neq 0$, using the following formula given by Troyer {\it et al.} \cite{Troyer}:
\begin{equation}
C(T) \propto \left(\frac{\Delta}{k_\mathrm{B}T}\right)^{3/2}\exp{\left(-\frac{\Delta}{k_\mathrm{B}T}\right)}.
\label{r}
\end{equation}
The values of $\Delta$ obtained by fitting eq. (6) to the low-temperature data are plotted in Fig. 3.
The field dependence of the gap is described well by the power law of eq. (1) with $A=5.19(2)$ and $\alpha=0.67(5)$, where $\Delta$ and $H$ are measured in K and T, respectively. The experimental exponent ${\alpha}$ is in good agreement with ${\alpha}=2/3$ derived from the quantum Sine-Gordon model \cite{Oshikawa2,Affleck}. By using eq. (3) with $J/k_{\rm B}\simeq 100$ K and $A=5.19$, the proportional coefficient $c_{\rm s}$ is evaluated as $c_{\rm s}\simeq 0.31$, which is smaller than $c_{\rm s}\simeq 0.54$ obtained from the susceptibility data.
\begin{figure}[ht]
\begin{minipage}{18pc}
\includegraphics[width=18pc]{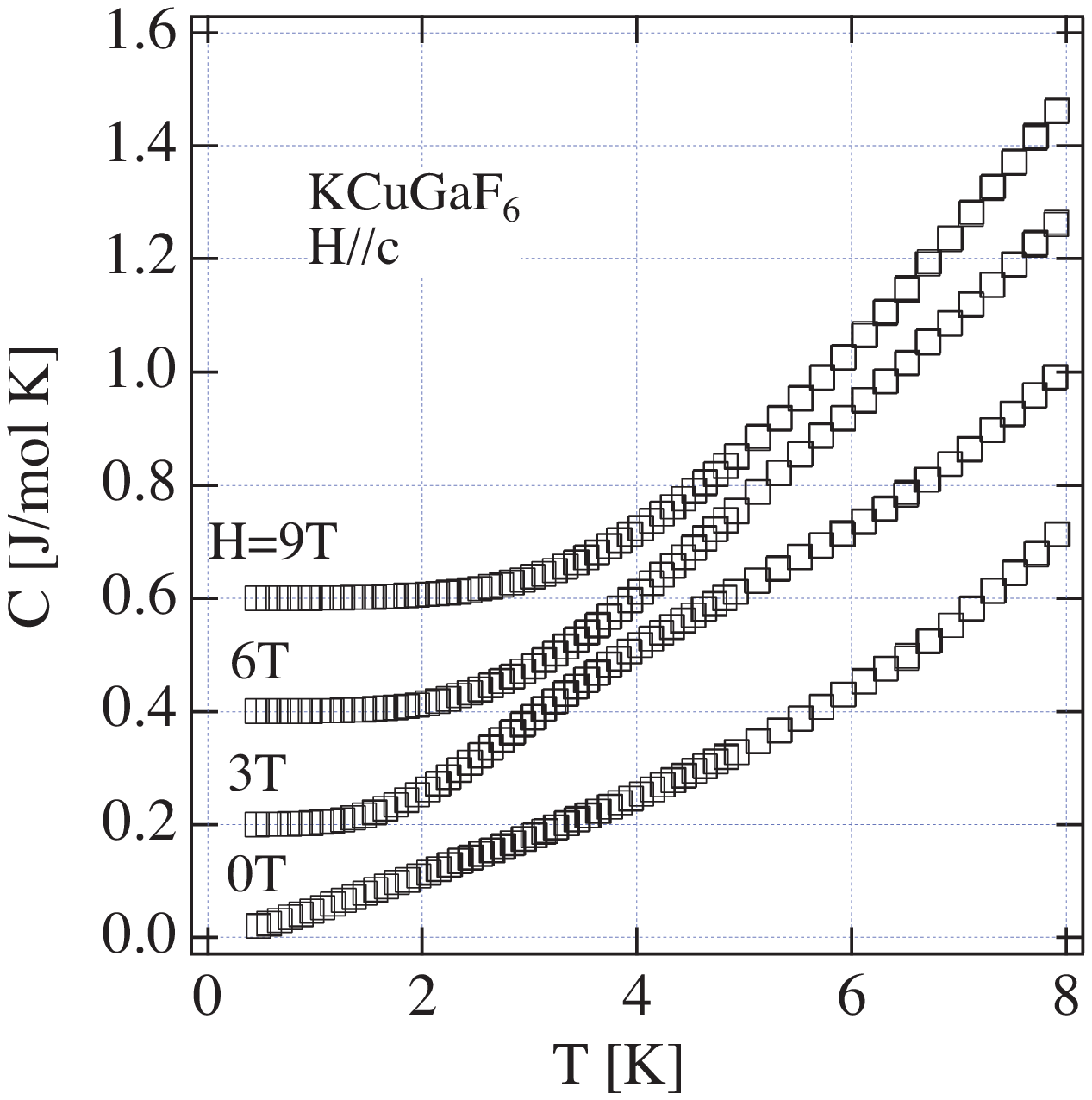}
\caption{Specific heat $C$ in KCuGaF$_6$ at various magnetic fields for $H\parallel c$. The values of  $C$ are shifted upward consecutively by 0.2 J/mol K. \label{bi}}
\end{minipage}\hspace{2pc}%
\begin{minipage}{18pc}
\includegraphics[width=17.6pc]{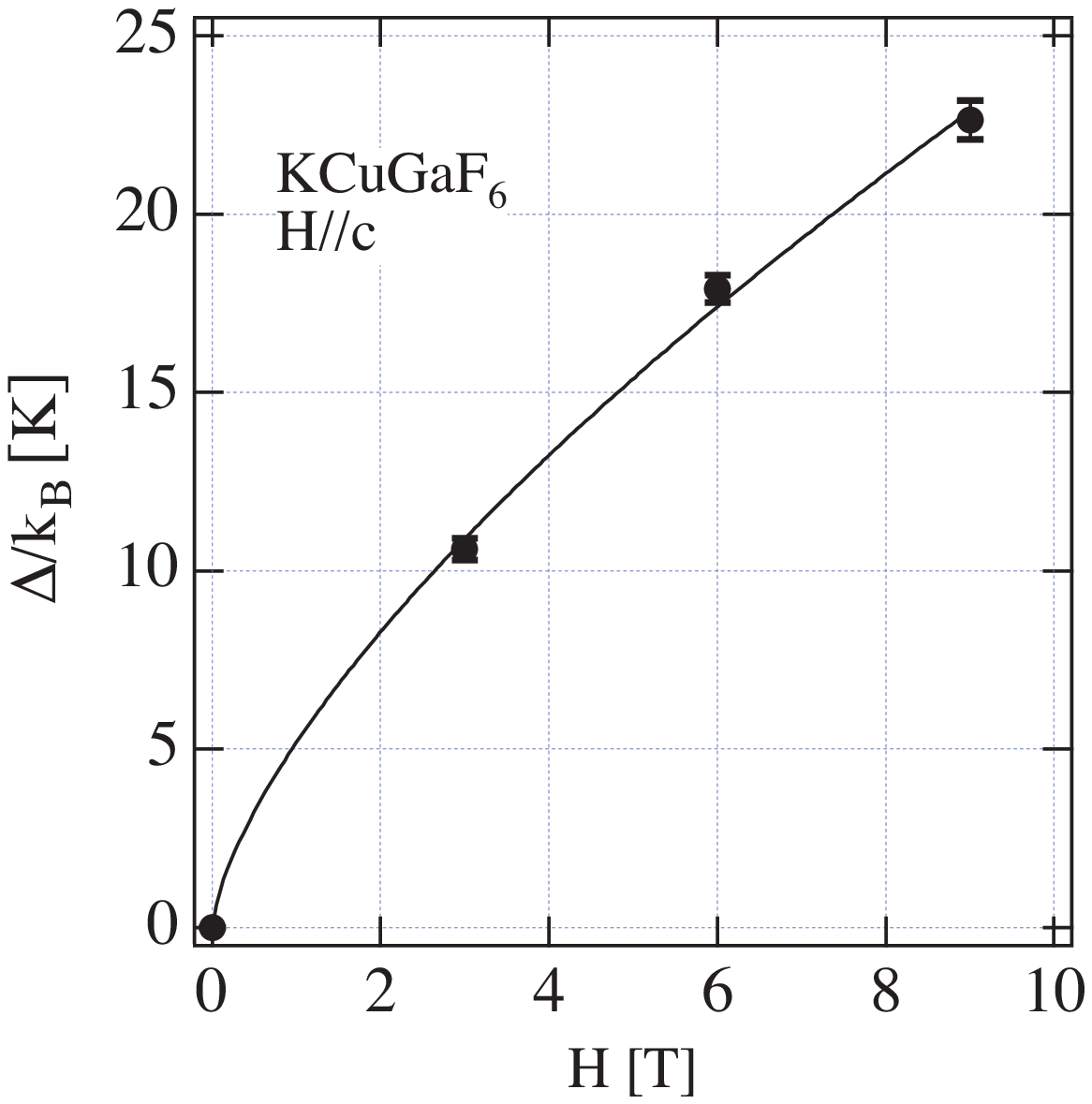}
\caption{ Field dependence of the gap in KCuGaF$_6$ obtained for $H\parallel c$. The solid line denotes the fit obtained using eq. (1) with $A=5.19$ and ${\alpha}=0.67$. \label{ci}}
\end{minipage} 
\end{figure}

\section{Conclusion}
We have presented the results of magnetic susceptibility and specific heat measurements performed on the $S=1/2$ antiferromagnetic Heisenberg chain system KCuGaF$_6$. We observed the Curie term in the susceptibility and the field-induced gap, both of which depend strongly on the field direction. These phenomena can be attributed to the effective staggered field induced by the applied magnetic field through the DM interaction and/or the alternating $g$-tensor. In large exchange interaction $J/k_{\rm B}\simeq 100$ K and large proportional coefficient, e.g., $c_{\rm s}\simeq0.31\sim 0.54$ for $H\parallel c$, KCuGaF$_6$ differs from other systems which exhibit the field-induced gap.

\ack
This work was supported by a Grant-in-Aid for Scientific Research and the 21st Century COE Program at Tokyo Tech ``Nanometer-Scale Quantum Physics'', both from the Ministry of Education, Culture, Sports, Science and Technology of Japan.

\end{document}